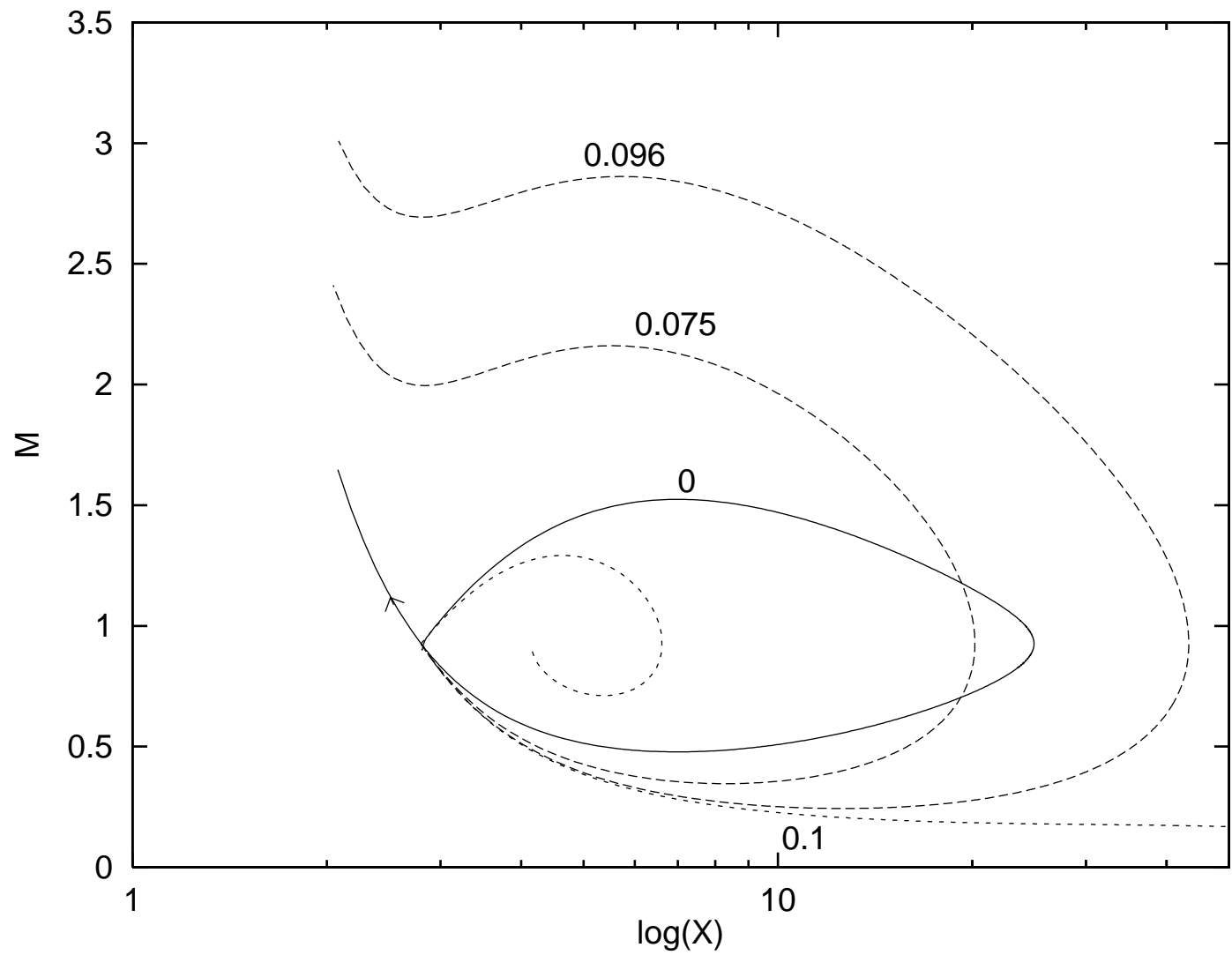

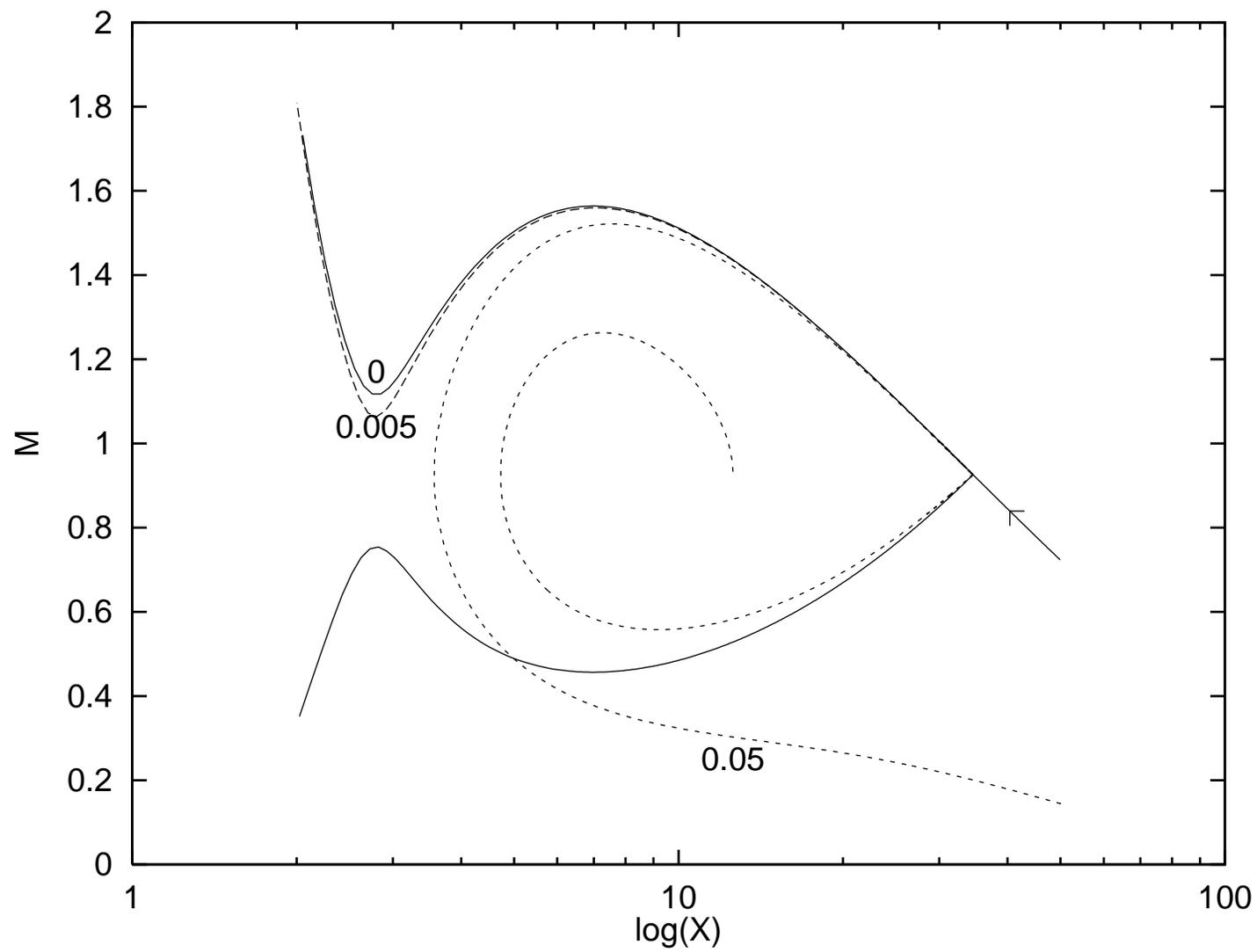

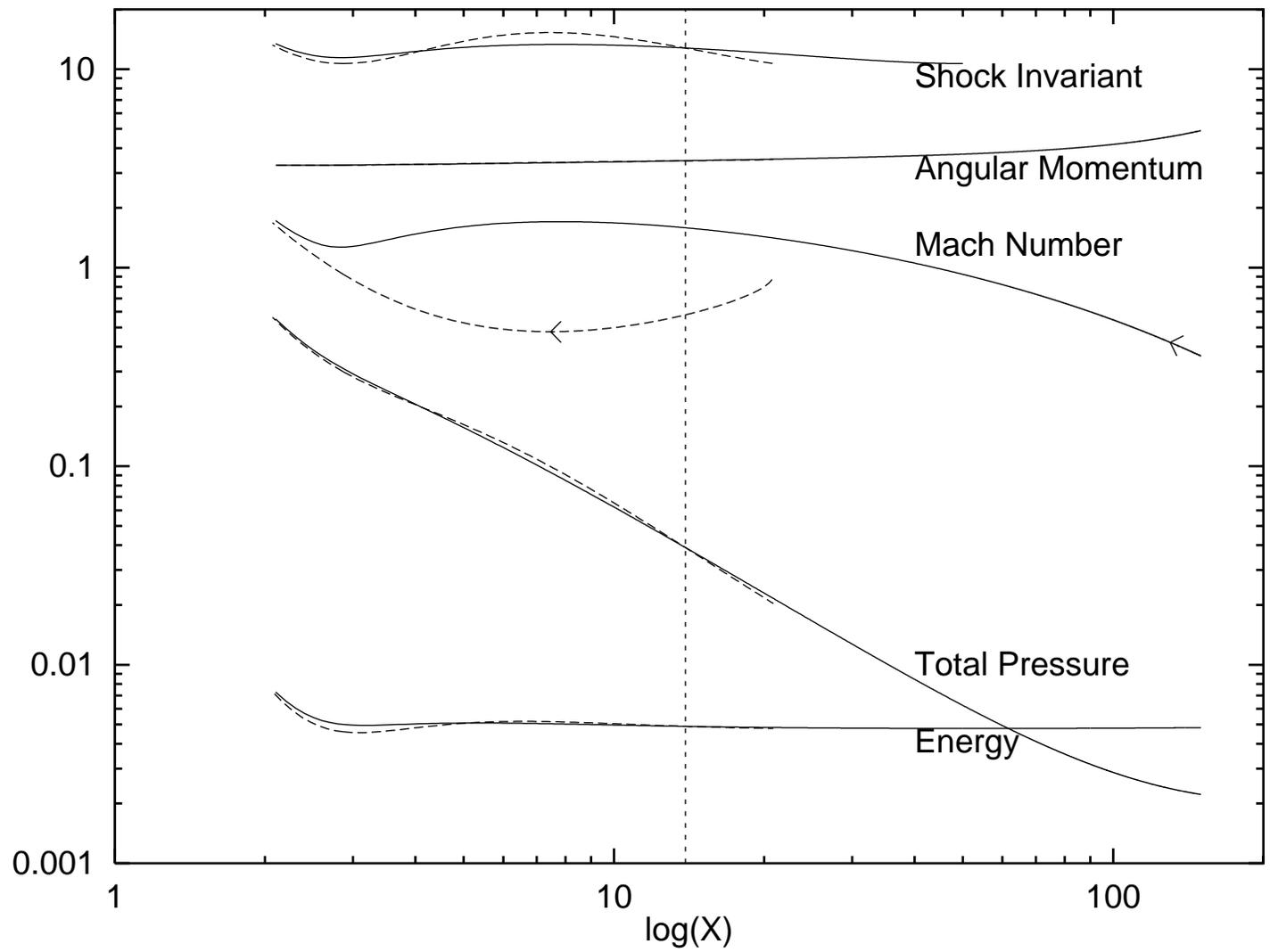

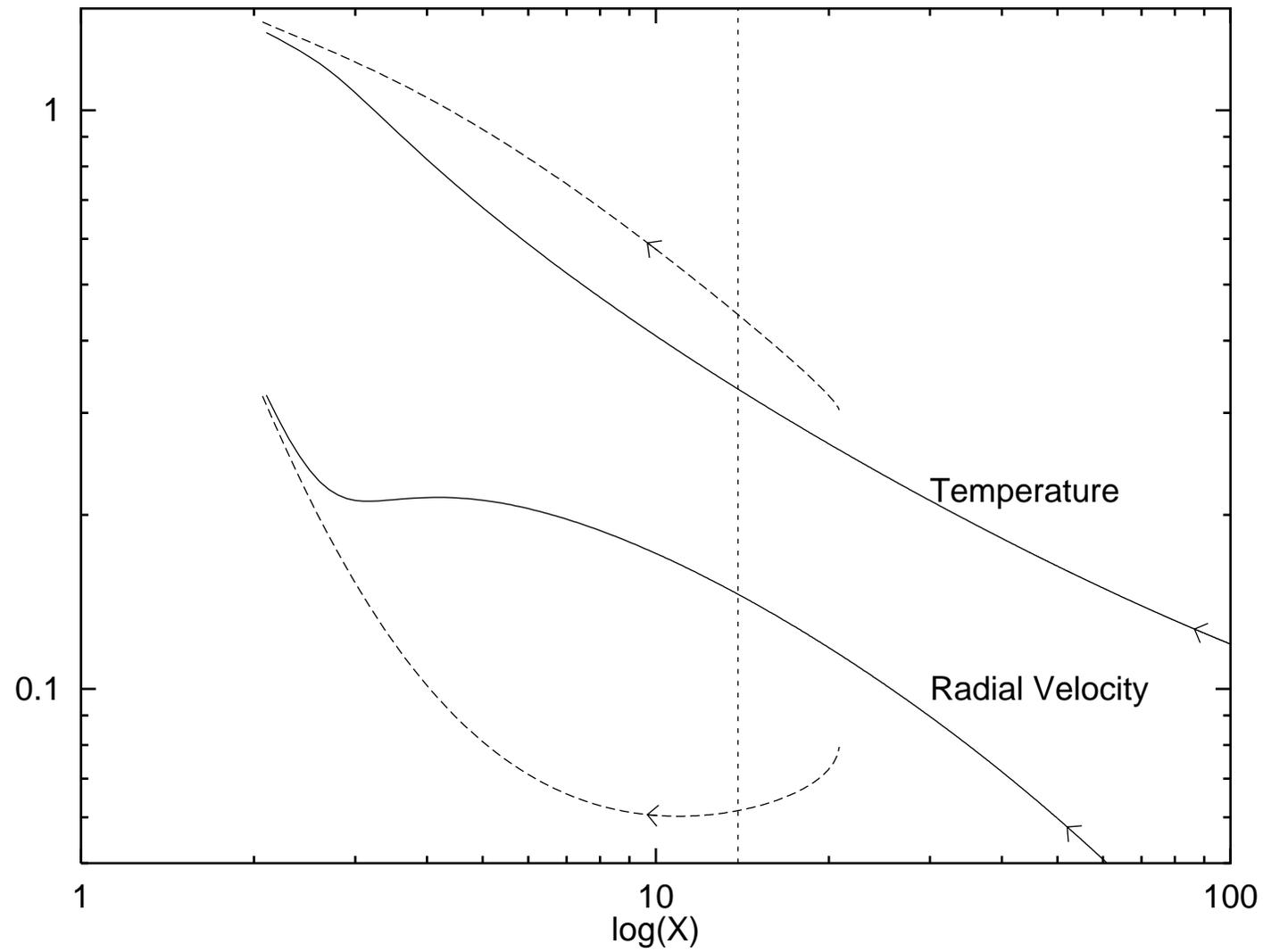

# Comments on the Newly Discovered Advection Dominated Flows Around Black holes and Neutron stars

Sandip K. Chakrabarti[1]

Goddard Space Flight Center, Greenbelt MD, 20771

e-mail: I: chakraba@twinkie.gsfc.nasa.gov

– 2 –

## ABSTRACT

We provide complete and global solutions of viscous transonic flows around black holes. We show that for any degree of advection, there are two critical viscosity parameters $\alpha_{c1,c2}$ such that for $\alpha < \alpha_{c1}$ the flow may pass through the inner sonic point only. For $\alpha_{c1} < \alpha < \alpha_{c2}$, the flow may have shock waves, and for $\alpha > \alpha_{c2}$, the flow may again pass through the inner sonic point, depending on flow parameters. No new topologies emerge other than what we found earlier while studying viscous isothermal transonic flows. These findings corroborate our unification scheme discussed in the context of the stable, transonic solutions around black holes and neutron stars.

[1] NRC Senior Research Associate

## 1. Introduction

The study of viscous, transonic flows was initiated by Paczyński & Bisnovatyi-Kogan (1981) to incorporate proper treatment of accretion disks close to a black hole, where radial velocity is significant, and where the standard Shakura-Sunyaev (1973) disks are likely to be unstable. This model was followed up by Abramowicz et al. (1988) where local computation of $\dot{m}(\Sigma)$ including advective flux showed stabilizing effect on the inner region of the Shakura-Sunyaev disks. First global solution of the viscous, transonic flows were obtained by (Chakrabarti, 1990ab; hereafter C90ab). For simplicity, these works assumed that the heating and cooling were adjusted in a manner that the temperature of the disk remained constant. In the language of Shakura-Sunyaev (1973) viscosity parameter $\alpha$, we showed that if $\alpha < \alpha_{cr}$, the incoming flow may either have a continuous solution passing through outer sonic point, or, it can have standing shock waves at $x_{s3}$ or $x_{s2}$ (following notations of C90ab). For $\alpha > \alpha_{cr}$, a shock wave at $x_{s2}$ persisted, but the flow now had two continuous solutions — one passed through the inner sonic point, and the other through the outer sonic point. Later analytical and numerical works (Chakrabarti & Molteni, 1993, 1995; hereafter CM93 and CM95 respectively) showed that $x_{s3}$ is stable, and that for $\alpha > \alpha_{cr}$ the continuous solution passing through the inner sonic point is chosen. We noted that $\alpha_{cr}$ ($\sim 0.015$ for the case considered) was a function of the model parameters, such as the disk temperature, sonic point location and angular momentum at the inner sonic point. Our result did not agree with the findings of Muchotrzeb (1983), who also studied isothermal flows and found that solutions should not exist for some $\alpha > \alpha_* \sim 0.02$.

Subsequently, extensive numerical simulations of quasi-spherical adiabatic accretion flows (Molteni, Lanzafame & Chakrabarti, 1994), showed that shocks form very close to the location where vertically averaged model of adiabatic flows predict them (Chakrabarti, 1989; hereafter C89). The flow advected its entire energy to the black hole and the entropy generated at the shock is also totally advected allowing the flow to pass through the inner sonic point. It was also found, exactly as predicted in C89, that flows with positive energy and higher entropy form strong, supersonic winds. In presence of viscosity also, very little energy radiates away (e.g., Fig. 8 of C90a). Having satisfied ourselves by the stability of these solutions (CM93, MLC94, CM95), we proposed an unified scheme of accretion disks (Chakrabarti, 1993, 1994, 1995; hereafter C939495, CM95) which combines the physics of formation of sub-Keplerian disks with and without shock waves depending on viscosity parameters and accretion rates. We always considered only the stable branch of the transonic flow and our solutions remained equally valid for black hole and neutron star accretions as long as different inner boundary conditions are employed.

Our present interest to revisit the problem arose after we became aware of newly discovered advection dominated models of accretion flows around black holes (Narayan & Yi, 1994; hereafter NY94) and the 'unified' scheme of accretion disks (Chen et al., 1995; hereafter CALNY95) which use cooling mechanisms different from C90ab. These solutions supposedly discuss new branches which become advection dominated close to the black holes. These solutions did not find shock waves (which were found to be very robust in simulations of CM93, CM95, MLC94, Ryu et al, 1995). Since no self-similar model may have any preferred length scales including sonic points and shock waves, shocks are not expected in NY94 treatment. Similarly, local solutions with the assumption of Keplerian angular momentum distribution $\Omega = \Omega_{Keplerian}$ as in CALNY95 could not have shock waves either, as the flow does not encounter any centrifugal barrier that is needed to form a shock wave (unless the flow is pre-heated by external radiations as in spherical accretion flows of Chang & Ostriker, 1985).

In this *Letter*, we present exact global solutions of the equations used by Narayan & Yi (1994), (but applicable to black hole accretions) and show that no new topology of solutions emerge other than what we had already obtained in the context of viscous, transonic, isothermal flows (C90ab). We neither assume the flow to be self-similar as in NY94 (and therefore Mach number of our disk is not a constant), nor assume the flow to be Keplerian as in CALNY95. As a result, as in isothermal flows (C90ab), we obtain solutions which allowed the disk to have shock waves as well. We also obtain a critical $\alpha$ parameter with exactly the same meaning as before. We therefore suspect that the so-called advection dominated flows do not constitute any new solutions. Our suspicion becomes stronger when we find that the conclusions of NY94 work regarding the advection of energy and entropy, formation of winds with positive energy and similarity of quasi-spherical results with those of vertically averaged model are exactly same as what we already discovered in references cited above.

## 2. Model Equations

We assume units of length, velocity and time to be $x_g = 2GM_{BH}/c^2$, $c$ and $2GM_{BH}/c^3$ respectively. In order to obtain complete and global topologies of the 'advection dominated' flows, we employ here the same set of equations as in NY94.

(a) The radial momentum equation:

$$\vartheta \frac{d\vartheta}{dx} + \frac{1}{\rho}\frac{dP}{dx} + \frac{l_{Kep}^2 - l^2}{x^3} = 0, \tag{1a}$$

(b) The continuity equation:

$$\frac{d}{dx}(\Sigma x \vartheta) = 0, \tag{1b}$$

(c) The azimuthal momentum equation:

$$\vartheta \frac{dl(x)}{dx} - \frac{1}{\Sigma x}\frac{d}{dx}(x^2 W_{x\phi}) = 0, \qquad (1c)$$

(d) The entropy equation:

$$\Sigma \vartheta T \frac{ds}{dx} = Q^+ - Q^- = f(x)Q^+, \qquad (1d)$$

Here $l_{Kep}$ is the Keplerian angular momentum, $W$, $\Sigma$ and $W_{x\phi}$ are the vertically integrated density $\rho$ and viscous stress respectively, $h(x)$ is the half-thickness of the disk at radial distance $x$ obtained from vertical equilibrium assumption (C89), $\vartheta$ is the radial velocity, $s$ is the entropy density of the flow, $Q^+$ and $Q^-$ are the heat gained and lost by the flow. Unlike NY94, we use Paczyński-Wiita (1980) potential to describe the black hole geometry. We also drop self-similarity assumption (unlike Chakrabarti, 1990c, where self-similarity in vertically averaged flows were used for the first time in studying nox-axisymmetric flows). We employ Shakura-Sunyaev (1973) viscosity prescription $W_{x\phi} = \alpha W$, where $W$ is the integrated pressure $P$. As noted earlier (CM95), it is probably more appropriate to use $W_{x\phi} \propto \Pi$ while studying shock waves since the total pressure $\Pi = W + \Sigma \vartheta^2$ is continuous across the shock and such a $W_{x\phi}$ keeps the angular momentum across of the shock to be continuous as well. Except for Eqn. 1d, other equations are the same as used in our previous studies (C89 and C90ab). In C89, Eqn. 1d was replaced by the adiabatic equation of state $P = K\rho^\gamma$ with entropy constant $K$ different in pre-shock and post-shock flows, and in C90a, Eqn. 1d was replaced by the isothermal equation of state, $W = K^2 \Sigma$ ($K$ being the sound speed of the gas). In the present *Letter*, for simplicity, we assume $f(x)$=const as in NY94. We have verified that our conclusions do not change when more general cooling laws are used instead. Details will be presented else where.

### 3. Results

We solve equations 1(a-d) using sonic point analysis (Liang & Thompson, 1980; C89, C90ab). Figs. 1(a-b) show variations of Mach number $M = \vartheta/a$, ($a^2 = \gamma P/\rho$, $\gamma$ being the polytropic index chosen to be 4/3 here) with the radial distance as $\alpha$ is varied. The angular momentum at the inner edge of the disk and the 'cooling parameter' $f$ are chosen to be $l_{edge} = 1.65$ (as in C89) and $f = 0.5$ for illustration. The Figures are drawn with sonic points at $x_c = 2.8$ in 1a and $x_c = 34.58$ in 1b respectively, exactly as in Fig. 7 of C89. For $\alpha = 0$, the solutions are represented by the solid curves (as in Fig. 7 of C89). In Fig. 1a, as $\alpha$ (marked on the curves) is increased, the closed topology opens up (long dashed curves) till $\alpha > \alpha_{c2} \sim 0.0975$ is reached when the flow completely opens up (short dashed curve) to connect with Keplerian disks at larger distance. The location at which this joining takes place depends on $\alpha > \alpha_{c2}$, provided other parameters ($x_c$, $l_{edge}$, $f$) are kept fixed. In Fig. 1b, as $\alpha$ is increased, open solutions (long dashed curve which could join Keplerian disks at a large distance) become closed (short dashed curve) when $\alpha > \alpha_{c2} \sim 0.01$. Other branch of the solution is also shown by a short dashed curve. Note that at the sonic point, $M \neq 1$, but a value appropriate for our one-and-a-half dimensional model (C89). An exact expression for $M(x_c)$ is: $M^2(x_c) = [\alpha^2 f + n + 1 + \{\alpha^2 f(\alpha^2 f + 1) + (n+1)^2\}^{1/2}]/[\gamma(2n+1)]$ where, $n = (\gamma - 1)^{-1}$. For $\alpha = 0$, this goes over to $M(x_c) = 2n/(2n+1)$ as in C89 and over to $M(x_c) = 1$ for $\gamma = 1$ (isothermal flow) as in C90a.

We note here that the above consideration of critical $\alpha_{c2}$ applies only if the flow parameters are already in the 'accretion shock' region (i.e., where the outer sonic point has less entropy that the inner sonic point, C89). If the flow parameters are in the 'wind region' (i.e., when the inner sonic point has less entropy that the outer sonic point, or if the outer sonic point does not exist at all), then the flow will continue to pass through the inner sonic point (and join with the Keplerian disk) untill anothe critical viscosity parameter $\alpha_{c1}$ is reached.

A solution may include a shock wave provided Rankine-Hugoniot conditions (or, any other generalized shock conditions, see, Abramowicz & Chakrabarti, 1980; C90b) are satisfied somewhere between the inner and the outer sonic points and the entropy at the inner sonic point is higher compared to that at the outer sonic point. *Maximum $\alpha$ for which shocks could form with inner sonic point at $x_c$ is $\alpha_{c2}$*. As noted earlier, $\alpha_{x_c}$ depends strongly on other parameters. For example, for $x_c = 2.8$, and $l_{edge} = 1.65$, $\alpha_{cr} = 0.01, 0.015, 0.075, 0.0975, 0.165$ for $f = 0, 0.1, 0.3, 0.5$, and $1.0$ respectively. Figs. 2(a-b) show an example of the solution of Eqns. 1(a-d) which includes a shock wave. These solutions were completely missed by NY94, Abramowicz et al. (1988) and CALNY95. Here, we choose $\alpha = 0.05$ and $f = 0.5$ for illustration, and employ $W_{x\phi} = -\alpha\Pi$ prescription to guarantee continuity of angular momentum through the shock. The solid curves represent the branch passing through the outer sonic point located at $x_{out} = x_c = 50$ and the long dashed curves represent the branch passing through the inner sonic point at $x_{in} = x_c = 2.8695$. Initially, the flow passes through $x_{out}$ and becomes supersonic. After a standing shock at $x_{s3} \sim 13.9$ (vertical short-dashed curve) it becomes subsonic and enters the black hole through $x_{in}$ along the long dashed curve. The flow chooses this solution and not that by solid curve for $x < x_{s3}$ since the entropy of the flow is higher at the inner sonic point (C89, CM93, MLC94). In Fig. 2(a), we plot the shock invariant (C89) function $C$, angular momentum distribution $l(x)$ ($\times 2$), the Mach number distribution $M(x)$, the total pressure $\Pi$ (in arbitrary units) and the local energy density $E(x) = 1/2\vartheta^2 + (\gamma - 1)^{-1}a^2 + 1/2l^2/x^2 - 1/2(x - 1)^{-1}$. The solid and dashed curves intersect at $x = x_{s3}$ consistent with a Rankine-Hugoniot condition assuming an infinitesimal shock thickness. In Fig. 2(b), we show how the temperature $T = \mu m_p a^2/\gamma k$ (in units of $10^{11}$K) and the radial velocity $\vartheta$ (in units of $c$) vary in a complete solution. Here, $\mu = 0.5$ for pure hydrogen, $m_p$ and $k$ are proton mass and Boltzman constant respectively. At the shock, temperature goes up and the velocity goes down, in the same way as in our earlier studies (C89, C90ab). We note here that for $\gamma > 1.5$ the vertically averaged flow does not have three sonic points and therefore shock solutions are possible only if the flow is already supersonic.

Since out transonic solutions with $f =$const have one *less* free parameter (as the flow satisfies sonic point conditions), the $\dot{M}(\Sigma)$ relation is monotonic with positive slope and represent the stable solutions. Whether or not a transonic solution is possible at all depends upon the nature of $d\vartheta/dx|_{x_c}(x_c, \alpha, f, l_{edge})$. Here, we have left out the topologies which produce winds (with and without shocks) but we have verified them to be similar to the isothermal solutions (C90ab). Details will be presented else where.

## 4. Discussion and Conclusions

In this *Letter*, we presented complete and global solutions of viscous, transonic equations which may or may not include shock waves, depending upon the accretion rate (determining the cooling rate), angular momentum at the inner edge as well as viscosity parameters. Though a new name ('advection dominated', NY94) has been recently coined to describe a few of these solutions, we find no new topologies other than those already present in the literature in the context of viscous, transonic, isothermal flows (C90ab). Unlike self-similar solutions of NY94, which allowed only constant Mach number flows, our solutions show very rich topological properties in the $M - X$ plane and also allow formation of shock waves.

Upon discovering no new topologies with different cooling laws, we are once again convinced of our unified scheme (C939495, CM95) for flows around a black hole. If the accretion rate and viscosity are such that a shock wave can form, the post-shock solution would behave as a thick accretion flows discussed earlier in the literature (e.g., Paczyński & Wiita, 1980; Rees et al., 1982) only more consistent, as advection is included as well. If the viscosity is high enough, the flow becomes so-called 'slim' disk or 'advection-dominated' disk, only more self-consistent, as we neither assume self-similar flows (NY94) nor locally Keplerian flows (CALNY95). Furthermore, our angular momentum distribution does not show inexplicable undulations seen in 'slim' disks (Abramowicz et al, 1988), but smoothly joins with Keplerian, even when shock waves are present. We show that for a given cooling parameter (e.g., $f$), a solution can match with Keplerian flows at larger distance, either when $\alpha < \alpha_{c1}$ or if $\alpha > \alpha_{c2}$ when the flow had three sonic points. If $\alpha_{c1}\alpha < \alpha_{c2}$, the solution includes a shock wave. We know of no other solutions in the Literature which are as exact and complete as we discussed in C90ab, and in this *Letter*. Just as C89 and C90ab solutions were found to be stable through time-dependent numerical simulations (CM93, MLC94, CM95), we believe that our present solutions containing shocks are also stable. Indeed, we have already verified that disks with a power law (in $T$) cooling (which includes bremsstrahlung) have stable shock waves (Molteni, Sponholz & Chakrabarti, 1995). These shocks show oscillations (with $5 - 10$ percent of amplitude in emitted radiations) whenever the cooling time roughly agrees with the in fall time and have all the characteristics of Quasi-Periodic Oscillations. Similarly, diversion of flows through the outer and inner sonic points mediated by viscosity (CM95) may be responsible for dwarf novae type outbursts. In another development, Chakrabarti & Titarchuk (1995) show that Comptonization of the soft photons (coming from the pre-shock Keplerian disks) in the hot post-shock region produces hard X-rays and $\gamma$-rays compatible with the observations of galactic and extra-galactic black hole candidates.

Although, we have used pseudo-Newtonian potential, the result may be exactly similar when full general relativity is used. For instance, inviscid flows in Kerr geometry (Chakrabarti 1990d) show no new topology other than what is in C89. Even fully general relativistic studies of magnatohydrodynamic flows (Takahashi et al, 1990, Englemaier, 1993) show no new topologies other than what was obtained with psudo-Newtonian geometry (Chakrabarti, 1990e). We therefore believe that our description of black hole and neutron star accretion is complete.

– 13 –

---

Fig. 1.— (a-b): Variation of Mach number with logarithmic radial distance as viscosity parameter $\alpha$ is varied. The advection parameter $f = 0.5$ is chosen. Solid curves ($\alpha = 0$) represent solution of C89. At the inner sonic point (a), for $\alpha < \alpha_{cr} \sim 0.0975$, and at the outer sonic point (b), for $\alpha < \alpha_{cr} \sim 0.01$ (solid and long dashed curves), the topologies are useful for shock formation. For $\alpha > \alpha_{cr}$, (short dashed curve) the topologies are open and useful for complete global solutions in (a). In (b), the topology becomes closed and this branch is not available for a black hole accretion.

Fig. 2.— (a-b): An example of a complete solution which includes a shock wave. $\alpha = 0.05$ and $f = 0.5$ are used. Solid curves show quantities as matter passes from subsonic to supersonic at $x_{out} = 50$. Shock conditions are satisfied at $x = 13.9$ (vertical dash curve) and the flow subsequently enters the black hole along the long dashed curves becoming supersonic at $x_{in} = 2.8695$. See text for details.